# Normal dispersion silicon oxynitride microresonator Kerr frequency combs

*Dongyu Chen[1], Andre Kovach[2], Sumiko Poust[3], Vincent Gambin[3], Andrea M. Armani[1,2,*]*

[1]Ming Hsieh Department of Electrical Engineering-Electrophysics, University of Southern California, Los Angeles, California 90089, United States

[2]Mork Family Department of Chemical Engineering, University of Southern California, Los Angeles, California 90089, United States

[3]NG Next, Northrop Grumman Corporation, Redondo Beach, California 90278, United States

**ABSTRACT**

On-chip optical resonators have proven to be a promising platform for generating Kerr frequency combs. Whispering gallery mode resonators are particularly attractive because of their small footprint as well as low threshold and power consumption. This performance can be attributed to two characteristics: the cavity quality factor (Q) and the cavity dispersion. The input optical field into the cavity is amplified by the cavity Q, enabling nonlinear processes to occur with low input powers. In addition, the total span of the optical comb is governed by the dispersion. In an optical cavity-based comb, the dispersion is governed by the geometric dispersion of the cavity and the material dispersion. While many different materials have been explored, the realization of ultra-high Q (UHQ) on-chip frequency combs sources is still challenging for most materials. One exception is the family of integrated silica devices. However, because the silica attracts water molecules from the ambient environment, the lifetime of the device performance is fundamentally limited, unless packaged in a controlled atmosphere. Here, we demonstrate the generation of environmentally-stable frequency combs fabricated from $SiO_{1.7}N_{0.13}$ microtoroidal resonators. Due to the UHQ factors of the cavities, parametric oscillations with sub-microwatt thresholds are demonstrated. Based on their geometry and material properties, the cavities have normal dispersion. However, due to avoided mode crossing, frequency combs are generated. Finally, unlike $SiO_2$, the $SiO_{1.7}N_{0.13}$ inhibits the formation of a water monolayer, allowing sub-microwatt performance to be maintained for over a week in devices stored with minimal environmental controls.

The development of novel device platforms for optical cavity-based Kerr frequency combs has attracted significant attention because of their small footprint, low loss, and high wavelength precision. Since the initial demonstration using silica microtoroids[1], on-chip Kerr frequency combs have been realized in many different materials including silicon[2], silicon nitride ($Si_3N_4$)[3-6], aluminum nitride (AlN)[7-9], aluminum gallium arsenide (AlGaAs)[10], lithium niobite[11], and diamond[12] as well as continued efforts in silica[13-16]. This diversity has enabled advances in telecommunications[17, 18], spectroscopy[19, 20], waveform generations[21], and many other applications[22, 23].

Among the different parameters governing comb generation, the cavity quality factor (Q) and the cavity dispersion are two of the most important parameters when considering a new device design for frequency comb generation. The threshold of the pump power is inversely proportional to the square of the quality factor[8, 12, 24], and the comb span is governed by the dispersion[3, 25]. Minimizing the pump power is essential for building integrated frequency comb chips with embedded laser sources, and many applications require large spanning combs. The primary limitations on an integrated device Q are related to either surface roughness from the device fabrication process or the optical loss of the material[26]. Despite great efforts devoted to increasing quality factors, the majority of the integrated devices have Q's below 100 million, thus fundamentally complicating integration efforts. To date, only silica devices have demonstrated frequency comb generation using an integrated device with a Q over 100 million[1, 15].

However, the Q factor of a silica device is transient. An intrinsic property of silica is that the surface consists of a monolayer of hydroxyl groups which attract water from the ambient environment[27, 28]. This water layer is extremely optically absorbing, decreasing the optical Q.

Therefore, as the water monolayer forms on the surface, the performance of the silica devices degrades, increasing the comb threshold. For frequency combs generated from silica devices, complex experimental controls are required to protect the devices from moisture, including performing measurements in nitrogen-purged boxes[1, 29] or building hermetic covers to protect the devices[15].

While the cavity Q controls the threshold, the cavity dispersion governs the comb span[1]. Specifically, the dispersion ($D_2$) is directly related to the resonant frequencies of the cavity through the expression:

$$D_2 = \omega_{m+1} - \omega_m - (\omega_m - \omega_{m-1}) \qquad (1)$$

where $\omega_m$ is the angular frequency of the resonance and m is the azimuthal mode number. $D_2/2\pi$ denotes the difference between the adjacent free spectral range (FSR) values, which is commonly used for representing the dispersion of the resonant cavities. Because dispersion is determined by the resonant cavity optical modes, it has two contributions: geometric dispersion and material dispersion. This balance is both a strength and a weakness of optical cavity resonators. While it provides a path to tune the dispersion by tailoring the cavity geometry, it also means that small changes in the material properties, specifically in the refractive index over the cavity's free spectral range (FSR), will change the dispersion. As a result, to have a stable optical frequency comb, it is not enough to maintain the quality factor of the device. The geometric dispersion and material dispersion must also be environmentally stable. Thus, new material systems and device platforms that can meet these demands are needed.

An alternative material to $SiO_2$ is $SiO_{1.7}N_{0.13}$. In previous research, it has been shown that the density of the hydroxyl groups on the surface of the $SiO_{1.7}N_{0.13}$ toroidal devices is much lower than silica devices, allowing the $SiO_{1.7}N_{0.13}$ devices to maintain ultra-high quality factors over weeks in ambient conditions[30]. However, the material properties, including the material dispersion, has not been measured. In this paper, frequency combs are generated from the $SiO_{1.7}N_{0.13}$ microtoroids (Figure 1) with a quality factor of $1.3\times10^8$, and the environmental stability of the frequency combs is demonstrated. Sub-microwatt thresholds are achieved due to the ultra-high Q of the device. The comb spectrum is measured and shown to be stable over time, even when stored in ambient environments due to the Q and dispersion stability.

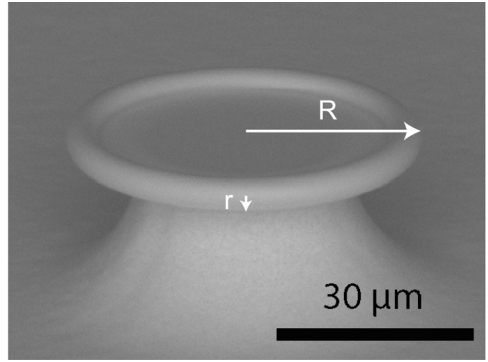

**Figure 1.** SEM image of a $SiO_{1.7}N_{0.13}$ microtoroid. R is the major radius of the toroid, and r is the minor radius of the toroid.

To determine the material dispersion of the custom $SiO_{1.7}N_{0.13}$ used, the refractive index of the $SiO_{1.7}N_{0.13}$ layer is measured using spectroscopic ellipsometry, and the results are shown in Figure 2(a). As expected, the refractive index of the $SiO_{1.7}N_{0.13}$ is slightly higher than silica because of the addition of nitrogen atoms into the material. Using these results, the material dispersion of the $SiO_{1.7}N_{0.13}$ layer can be calculated using

$$D = -\frac{2\pi c}{\lambda^2}\frac{d^2\beta}{d\omega^2} \quad (2)$$

where c is the speed of light, $\lambda$ is the wavelength, $\beta$ is the propagation constant, and $\omega$ is the angular frequency (Figure 2(a)). Based on these measurements and calculation, the zero-dispersion point for the $SiO_{1.7}N_{0.13}$ is determined to be around 1270 nm. In comparison to silica, this point is shifted by 100 nm to the red.

As discussed, the dispersion of the microtoroid device is a combination of both the material dispersion and geometric dispersion[1]. The geometric dispersion of the cavity is influenced by the major (R) and minor (r) radius of the toroids. To calculate the overall dispersion, simulations are performed using COMSOL Multiphysics. The resonant frequencies of the cavity modes are calculated using the software with the $SiO_{1.7}N_{0.13}$ material dispersion embedded. The dependence of the overall dispersion on the major/minor radius for a series of device geometries is investigated in Figure 2(b) and (c). In Figure 2(b), the major radius is held constant at 60 µm and the minor radius varies from 2.5 µm to 5 µm. In Figure 2(c), minor radius of the toroid is held constant at 2.5 µm and the major radius varies from 20 µm to 60 µm. These device geometries are chosen as they cover the device size that is experimentally studied in the present work (R=27 µm and r=2.83 µm).

The simulation results show that as the major radius increases and the minor radius decreases, the dispersion curves shift from normal dispersion to anomalous dispersion; thus, the zero-dispersion point shifts to a lower wavelength. For example, for a toroid with R=60 µm and r=2.5 µm, the zero-dispersion point is approximately at 1380 nm which is 200 nm lower than a device with r=5µm. However, reproducibly fabricating $SiO_{1.7}N_{0.13}$ toroidal devices with UHQ factors with a large major radius and small minor radius is very challenging because of the thermodynamic, self-limiting nature of the laser reflow process.

Given this complexity, we chose to use $SiO_{1.7}N_{0.13}$ toroids with normal dispersion in our experiment. The details of the device fabrication are in the supplementary material. The radii of the devices are measured to be R=27 µm and r=2.83 µm using both optical and electron microscopy. The dispersion curve of these devices is simulated and plotted in Figure 2(c) with the black dashed line. The dispersion of the device is also experimentally measured using swept-wavelength interferometry. The measured dispersion points are plotted in Figure 2(c), illustrated by stars. To more clearly inspect the results, only the indicated region of the data is plotted in Figure 2(d). The experimentally measured dispersion confirms the normal dispersion predicated by the simulation. Details of the dispersion measurement are included in the supplementary material.

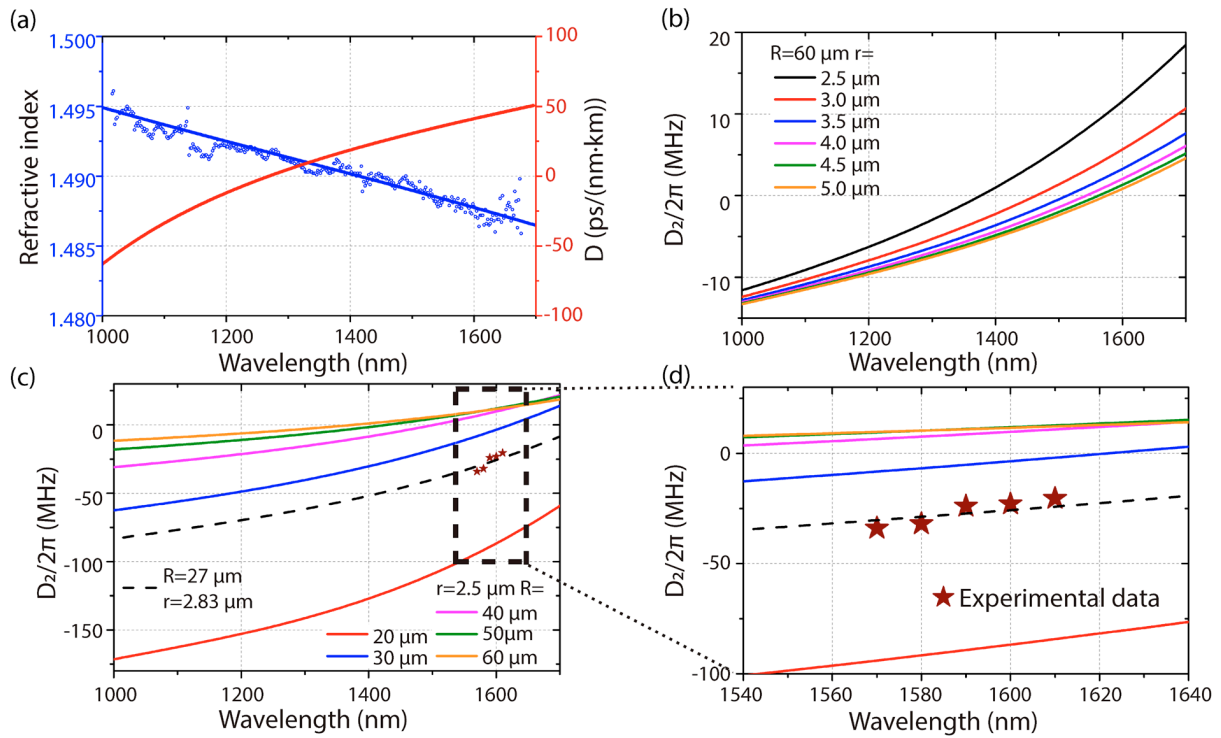

**Figure 2.** (a) Blue: Measured refractive index of the $SiO_{1.7}N_{0.13}$ layer. The results are fit to the Sellmeier equation. Red: Material dispersion of $SiO_{1.7}N_{0.13}$. (b) Dispersion of the microtoroids with R=60 μm and r=2.5, 3, 3.5, 4, 4.5, and 5 μm, respectively. (c) Solid: Dispersion of the microtoroids with r=2.5 μm and R=20, 30, 40, 50, and 60 μm, respectively. Dashed: Dispersion of the microtoroid with the same dimensions as used in the experiment (R=27 μm and r=2.83 μm). Stars: Experimentally measured dispersion values of the device. (d) Zoomed in simulated

dispersion curves and experimentally measured dispersion points in the 1540 nm – 1640 nm wavelength range.

Although anomalous dispersion is desired to achieve the phase matching conditions required for hyperparametric oscillation, it is challenging to achieve anomalous dispersion using $SiO_{1.7}N_{0.13}$ microtoroidal resonators as shown in Figure 2. Previous research has shown that, due to the avoided mode crossing between modes from different mode families, frequency combs can also be generated from cavities with normal dispersion[4, 31-34]. The interaction between modes alters the resonant frequencies of the modes, which can dramatically change the dispersion. Local anomalous dispersion can be achieved through this mode coupling. Thus, modulational instability is possible in microresonators with normal dispersion, such as the $SiO_{1.7}N_{0.13}$ toroids fabricated in this work.

One key feature of frequency comb generated through avoided mode crossing is that one initial sideband is fixed at a constant location when the pump laser is tuned into other modes in the same mode family. Two types of frequency combs can be generated depending on the relative position of the pump and the fixed initial sideband. Type-I frequency combs are generated when the initial sidebands are located one FSR away from the pump. In a Type-I comb, the cascaded four wave mixing (FWM) process duplicates the frequency spacing between the pump and the initial sidebands, leading to a series of comb lines spaced by one FSR. Type-II frequency combs are generated when the initial sidebands are located several FSRs away from the pump. The subsequent FWM process produces more comb lines to fill in the gap, including the lines with one FSR spacing[32, 34].

Frequency combs generated by $SiO_{1.7}N_{0.13}$ microtoroids with R=27 μm and r=2.83 μm are plotted in Figure 3. Details of the testing setups are described in the supplementary material. Figure 3(a) shows the full spectrum of a Type-I frequency comb. This comb has initial sidebands located one FSR away from the pump with a pump wavelength of 1551 nm. The 350 nm wide frequency comb is generated with ~60 mW pump power in a device with a Q of $1.3 \times 10^8$. The distance between two adjacent comb lines near the pump is around 9.55 nm, which agrees very well with the 9.53nm FSR predicated by the simulation.

To further study and verify the avoided mode crossing behavior, the pump laser is tuned from 1551 nm to 1560.6 nm and 1570.3 nm to excite the adjacent modes. The comb changes from type I to type II with initial sidebands located at two FSRs and three FSRs away from the pump, respectively, as shown in Figure 3(b). One initial sideband is fixed at approximately 1541.5 nm as indicated by the red stars which is clear evidence of the avoided mode crossing.

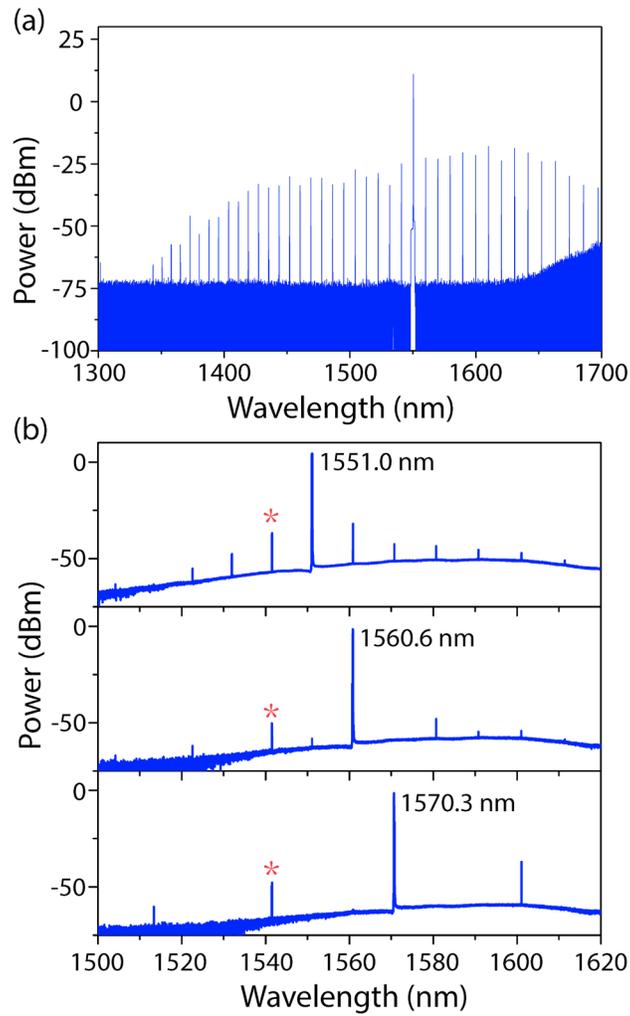

**Figure 3.** (a) Optical spectrum of a 350 nm wide frequency comb with a pump wavelength at 1551 nm and an input power of 60mW. (b) Comb generation at different pump wavelengths with the initial sideband fixed at 1541.5 nm as indicated by the red stars.

The threshold power required for the parametric oscillation is one of the most important parameters used to characterize comb generation. A low threshold is critical for decreasing the complexity and reducing the footprint of the system. The theoretical threshold ($P_{th}$) of the parametric oscillation is[8, 12]

$$P_{th} \approx 1.54 \frac{\pi}{2} \frac{Q_C}{2Q_L} \frac{n^2 V}{n_2 \lambda_P Q_L^2} \tag{3}$$

where $Q_C$ and $Q_L$ are the coupling and loaded quality factor. n is the refractive index, V is the mode volume. $n_2$ is the nonlinear refractive index, and $\lambda_P$ is the pump wavelength. Different strategies have been used to decrease the threshold of the parametric oscillation, such as increasing the intrinsic quality factors of the cavities, increasing the nonlinear refractive index, and decreasing the mode volume. In the present devices, the $SiO_{1.7}N_{0.13}$ microtoroids have very low threshold as a result of the high quality factors.

The power of the idler peak generated by the parametric oscillation is used to characterize the threshold of the process. As is shown in Figure 4, the power of the idler increases markedly with the input power coupled into the resonant cavity. The threshold of the parametric oscillation ($P_{th}$) is characterized to be around 280 µW. This threshold value is comparable to or lower than the threshold of most of the current on-chip devices used for Kerr frequency comb generation[8, 12, 35, 36]. For example, the threshold value for a $SiO_2$ microtoroid device with similar size and a Q of $5 \times 10^7$ is 339 µW[24], and the threshold values for large silica microdisks with Qs over $1.8 \times 10^8$ are several milliwatts[16]. However, while achieving a low threshold is important, the stability of the threshold is critical. If the value changes with time, the system's performance is unstable and unreliable; this poses a fundamental limitation to the comb's utility.

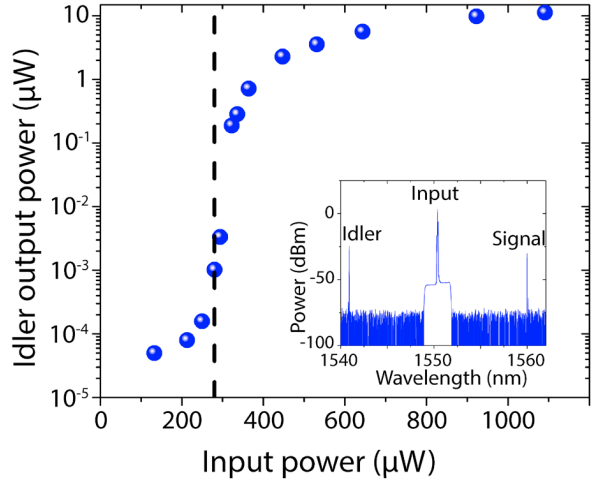

**Figure 4.** Power of the idler signal recorded on OSA as a function of the input power coupled into the resonator. The threshold is estimated to be around 280 µW.

One unique feature of $SiO_{1.7}N_{0.13}$ microtoroidal resonators is their ability to maintain high Q stability in ambient environments[30]. Given the relationship between Q and comb generation, this Q stability should result in an environmentally robust frequency comb generator with reproducible performance. However, while Q stability is required for reproducible comb performance, it is not sufficient as an isolated variable, because the comb generation process relies on more parameters than just Q, as can be observed from Equation 3. Therefore, demonstrating environmental stability represents a critical step forward in the field.

To investigate the stability of the $SiO_{1.7}N_{0.13}$ resonant combs, the frequency comb spectrum of the device is characterized both immediately after device fabrication and nine days later. The same mode is pumped for both measurements under the same coupling conditions. During the intermediate period, the device is stored in a gel-pak under ambient conditions (room temperature, atmosphere, and pressure). Based on previous work investigating water monolayer formation on silica surfaces[27, 30], this timeframe is sufficient for characterizing the potential change in device behavior. The Q factors recorded on day 1 and day 9 are both $1.3 \times 10^8$, indicating that the device Q does not change within the error of the measurement over the 9-day experiment duration.

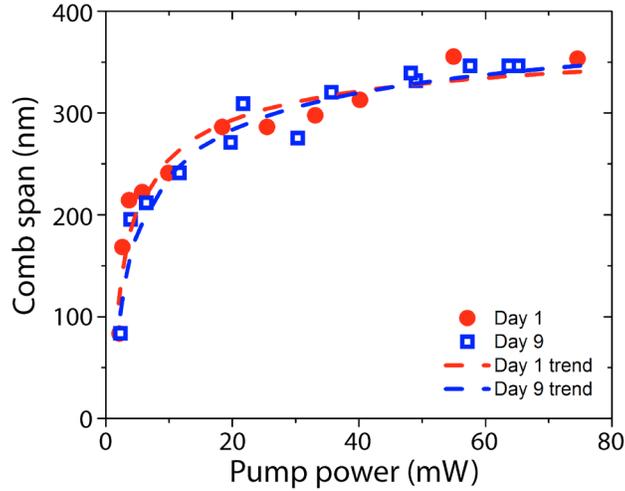

**Figure 5.** The relationships between the comb span and the input power measured at both Day 1 and Day 9 after the fabrication of the device. The combs are in Figure S4.

A pair of combs from day 1 and day 9 are presented in Figure S4 in the supplementary material. To quantify the similarity between the frequency combs, the relationship between the comb span and the pump power is characterized. The comb span is selected as an indicator because it not only depends on the quality factor of the device but also on the overall dispersion and all mode interactions. For devices working in the normal dispersion range, the comb span relies strongly on the optical mode distributions, which are very sensitive to surface and environmental changes. As such, comb span is more sensitive to environmental changes than threshold. Figure 5 shows the relationship between the comb span and the pump power for the frequency combs observed on these two different days. For both days, the comb span increases while the input power increases until it reaches its maximum. The two sets of measurements agree well with each other, indicating that both the Q and the dispersion of the device are unchanged even though the device has been left in an ambient atmosphere for 9 days.

In conclusion, we have demonstrated the generation of frequency combs from $SiO_{1.7}N_{0.13}$ microtoroids with normal dispersion. The dispersion of the microtoroid is studied with different major and minor radii. Both Type-I and Type-II frequency combs are demonstrated because of the avoided mode crossing. Due to the smooth cavity surface and the low material loss, the $SiO_{1.7}N_{0.13}$ devices show ultra-high quality factors, which are comparable to previous silica devices. These Q factors enable parametric oscillation to be achieved with threshold powers as low as 280 μW. Due to the environmental stability of the ultra-high-Q factors and the material dispersion of the $SiO_{1.7}N_{0.13}$ microtoroid resonators, they can be operated in ambient environments for several days with no detectable change in comb performance. This stability will greatly simplify integrated comb generators by relaxing the tolerances on packing requirements, paving the way for lighter weight atomic clocks, frequency converters, and spectroscopy systems[17-20, 23].

**Supplementary Material**

See supplementary material for the device fabrication and characterization details and additional experimental results.

**AUTHOR INFORMATION**

**Corresponding Author**

*E-mail: armani@usc.edu

**Author Contributions**

The manuscript was written through contributions of all authors. D. C. fabricated the devices, performed the simulation, characterized the devices, and analyzed data. A. K. assisted in device characterization. S. P. and V. G. fabricated the wafer. A. M. A. aided in experimental design and data analysis. All authors have given approval to the final version of the manuscript.

**Funding Sources**

Northrop Grumman Institute of Optical Nanomaterials and Nanophotonics and the Office of Naval Research (ONR) (N00014-17-2270).

*Supplementary Information*

# Normal dispersion silicon oxynitride microresonator Kerr frequency combs

*Dongyu Chen[1], Andre Kovach[2], Sumiko Poust[3], Vincent Gambin[3], Andrea M. Armani[1,2,\*]*

[1]Ming Hsieh Department of Electrical Engineering-Electrophysics, University of Southern California, Los Angeles, California 90089, United States

[2]Mork Family Department of Chemical Engineering, University of Southern California, Los Angeles, California 90089, United States

[3]NG Next, Northrop Grumman Corporation, Redondo Beach, California 90278, United States

**Device fabrication**

The $SiO_{1.7}N_{0.13}$ microtoroids are fabricated from a 1.5 μm thick $SiO_{1.7}N_{0.13}$ layer. The $SiO_{1.7}N_{0.13}$ layer is deposited on a silicon substrate using silane, ammonia, nitrogen, and nitrous oxide in a PECVD reactor at 250 °C and 900 mTorr. The RF power is 20 W. The compositional ratio is determined to be $SiO_{1.7}N_{0.13}$ using EDX. The device fabrication process comprises four steps.

First, photolithography is used to pattern 80 μm diameter circles on the wafer. Then, the $SiO_{1.7}N_{0.13}$ layer is etched using buffered oxide etchant, followed by $XeF_2$ etching to undercut the $SiO_{1.7}N_{0.13}$. Finally, a laser-assisted reflow is performed with a high-power $CO_2$ laser to smooth the surface of the device. After reflow, the major diameters of the devices are around 54 μm. Additional details on $SiO_{1.7}N_{0.13}$ layer characterization and device fabrication are contained in a previous work[1].

## Device characterization

Two different testing set-ups were used to characterize the devices (Figure S1 and S2). Figure S1 shows the testing setup used to experimentally verify the optical performance or Q of the $SiO_{1.7}N_{0.13}$ devices and measure the frequency comb spectra. A tunable laser working in the range of 1550 nm to 1630 nm (Newport Velocity Widely Tunable Laser, 1550nm-1630nm, TLB-6730) is amplified by an erbium-doped fiber amplifier (EDFA, Amonics, AEDFA-23-B-SA). Light is evanescently coupled into the microtoroid resonator through a tapered fiber waveguide (Newport F-SMF-28). A motion stage is used to control the coupling gap between the tapered fiber and the device. The transmission signal passes a 90:10 splitter and is detected simultaneously on both the optical spectrum analyzer (OSA, Yokogawa AQ6370C) (90%) and the photodetector (ThorLabs PDA10CS) (10%). The signal received by the photodetector is sent to an oscilloscope. The loaded quality factor is measured by continuously scanning the tunable laser over the resonant wavelength and by recording the resonant linewidth in the under-coupled regime. A coupled-cavity model is used to calculate the intrinsic Q from this linewidth. The comb spectra are recorded on the OSA. The threshold for parametric oscillation is determined by recording a series of spectra while varying the power coupled into the cavity.

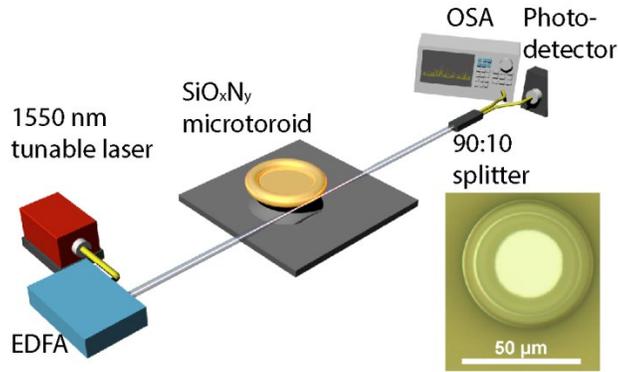

Figure S1. The schematic of the testing setup. Inset: a microscope image of the $SiO_{1.7}N_{0.13}$ microtoroid.

## Dispersion measurement

Dispersion of the device is characterized using swept-wavelength interferometry. Figure S2 shows the testing setup used in the experiment. A tunable laser (Newport Velocity Widely Tunable Laser, 1550nm-1630nm, TLB-6730) is split by a 50%/50% splitter. One laser signal is sent through a polarization controller and a tapered fiber to couple into the resonant cavity. The transmission is received by a photodetector (ThorLabs PDA10CS). The other laser signal is sent to a Mach-Zehnder interferometer (MZI). The two arms of the MZI have a 250-meter difference in length, which results in a ~0.8 MHz optical frequency sampling resolution. The MZI is calibrated using a standard gas cell (WavelengthReferences TRI-H(80)-5/150/150-FCAPC). The transmission signals from the MZI and the cavity are recorded by a two-channel data acquisition device (DAQ, National Instruments PCI-5114) simultaneously.

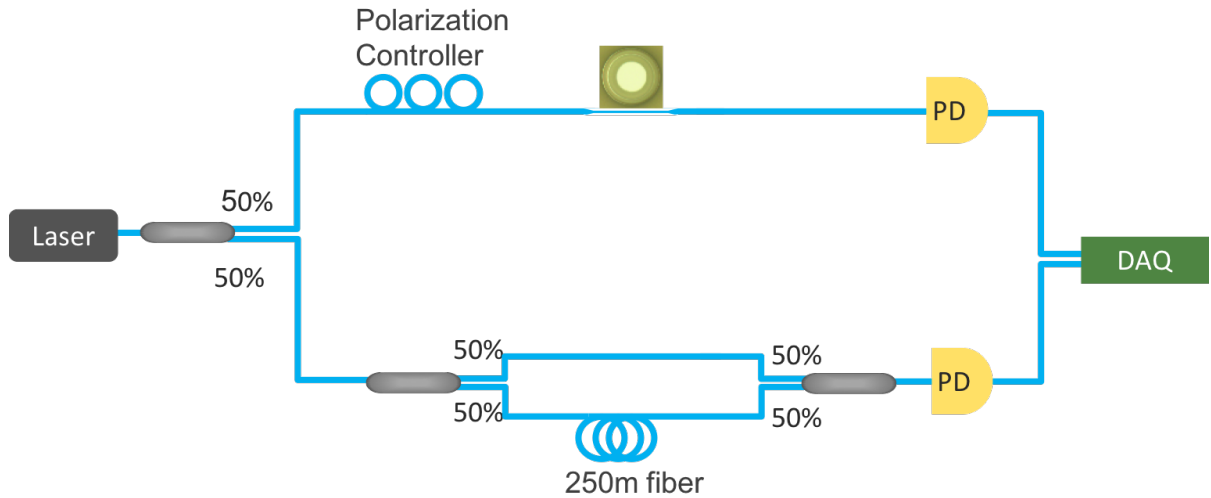

Figure S2. Testing setup used for the dispersion measurement.

When measuring the dispersion, the laser is scanned from a higher wavelength to a lower wavelength at a speed of 10nm/s to avoid thermal distortion of the optical modes. The calibration and measurement are separated into two wavelength ranges: 1630nm-1580nm and 1600nm-1550nm because of the memory limitation of the DAQ. Due to the large free spectral range (FSR) of our resonators, special attention is paid to the data analysis to lower the error level. The sampling signal from the MZI is statistically analyzed as well as manually examined to remove artifacts from laser scanning error (stage scanning jitter). Figure S3 shows the measured FSR of the resonator demonstrated in our experiments. These results are also plotted in Figure 2 in the main text to enable direct comparison with the theoretical calculations.

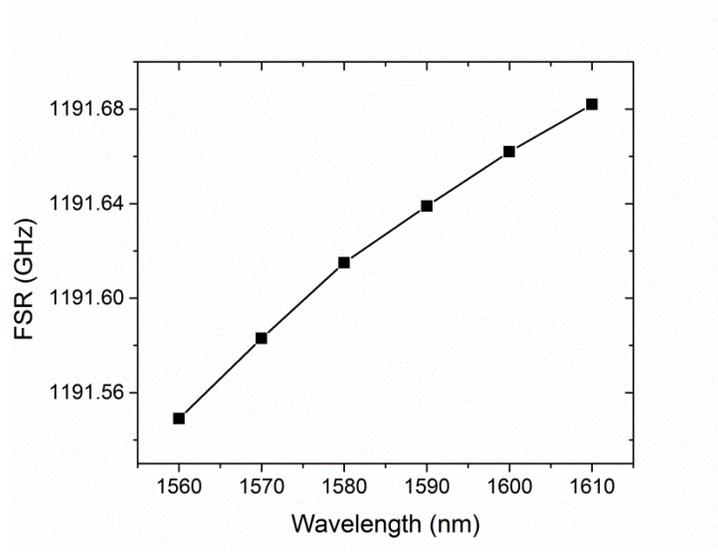

Figure S3. Measured FSR of the $SiO_{1.7}N_{0.13}$ microtoroid.

## Environmental Stability

Figure S4 shows a pair of recorded spectrum with the corresponding input power indicated in the upper-left corner. The comb spectrum at four different pump power levels are plotted, and similar comb spans and intensities are observed on day 1 (Figure S4a) and day 9 (Figure S4b). Notably, within the resolution of the measurement, the comb teeth are located at the same positions in these two sets of measurements.

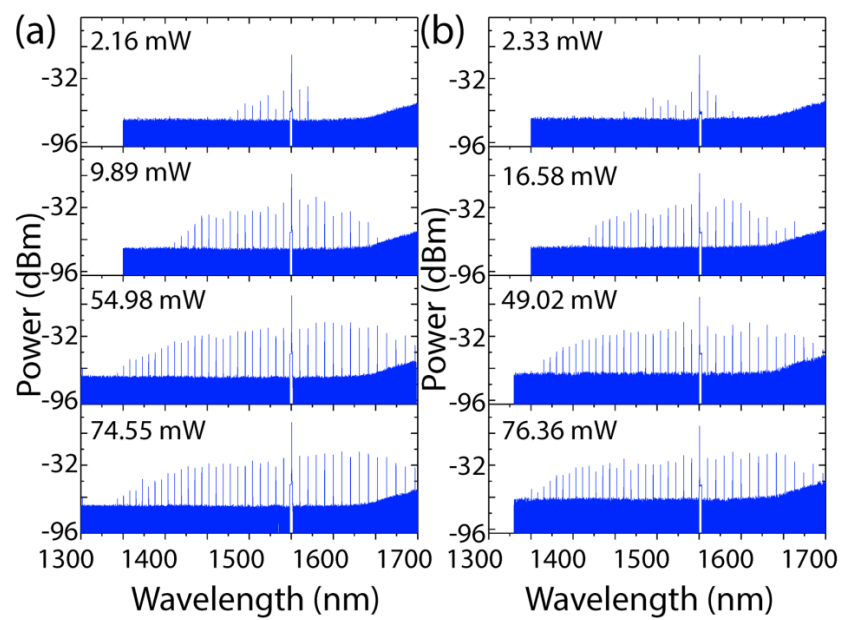

Figure S4. The spectrums of the frequency combs generated on (a) day 1, and (b) day 9 after the fabrication.